\documentclass[preprint,prl,aps,amsfonts,amssymb,amsmath,showpacs]{revtex4}
\usepackage{graphicx}% Include figure files
\usepackage{dcolumn}% Align table columns on decimal point

\graphicspath{{figures/}}
\setkeys{Gin}{width=\linewidth}

\begin{document}

\title{Scanning Gate Spectroscopy of transport across a Quantum Hall Nano-Island}

\author{F. Martins$^{1,*}$, S. Faniel$^{2}$, B. Rosenow$^{3}$, M. G. Pala$^{4}$, H. Sellier$^{5}$, S. Huant$^{5}$, L. Desplanque$^{6}$, X. Wallart$^{6}$, V. Bayot$^{1,5}$ and B. Hackens$^{1,*}$\\}

\address{$^{1}$ Universit\'e catholique de Louvain, IMCN/NAPS, Louvain-la-Neuve, Belgium\\
$^{2}$ Universit\'e catholique de Louvain, ICTEAM/ELEN, Louvain-la-Neuve, Belgium\\
$^{3}$ Institute for Theoretical Physics, Leipzig University, D-04009, Leipzig, Germany\\
$^{4}$ IMEP-LAHC, Grenoble INP, Minatec, BP 257, F-38016, Grenoble, France\\
$^{5}$ Institut N\'eel, CNRS and Universit\'e Joseph Fourier, BP 166, F-38042, Grenoble, France\\
$^{6}$ IEMN, Cit\'e scientifique, BP 60069, F-59652, Villeneuve d'Ascq, France\\
$^{*}$ Correspondence to: (F.M.) frederico.rodrigues@uclouvain.be, (B.H.) benoit.hackens@uclouvain.be.}

\begin{abstract}
We explore transport across an ultra-small Quantum Hall Island (QHI) formed by closed quantum Hall edge states and connected to propagating edge channels through tunnel barriers.
Scanning gate microscopy and scanning gate spectroscopy are used to first localize and then study a single QHI near a quantum point contact.
The presence of Coulomb diamonds in the spectroscopy confirms that Coulomb blockade governs transport across the QHI.
Varying the microscope tip bias as well as current bias across the device, we uncover the QHI discrete energy spectrum arising from electronic confinement and we extract estimates of the gradient of the confining potential and of the edge state velocity.

\end{abstract}


\maketitle

\section{Introduction}
Most peculiar properties of quantum Hall systems stem from the propagation of electrons along one-dimensional edge channels, emerging wherever a Landau level crosses the Fermi energy \cite{Halperin_PRB1982}. The so-called Edge States (ES), that mainly form at the borders of a two-dimensional sheet of electrons in a high perpendicular magnetic field $B$, provide a powerful model that describes the behavior of quantum Hall nano-devices~\cite{Buttiker_PRB1988}. Since scattering is topologically prohibited, ES offer a natural protection to electrons, so that they constitute ideal models of quantum wires, whose manipulation is particularly fruitful. Indeed, various device geometries can be carved from two-dimensional electron systems thanks to lithography techniques, and the position of ES within a device can then be tuned using voltages applied on metallic gates deposited on top of the device.

In this context, a strong interest recently emerged for interferometer geometries, where counterpropagating ES are brought close enough to interact, and electron loops can be formed at will. Tuning the coupling between ES leads to a large variety of situations.
In open systems, \emph{i.e.} when the conductance of the device $\sigma$ is much larger than $2e^2/h$, electronic analogs of Mach-Zehnder~\cite{Ji_Nat2003,Neder_PRL2006,Roulleau_PRL2008} and Fabry-P\'erot~\cite{Rosenow_PRL2007, McClure_prl2009,Zhang_PRB09,Yamauchi_PRB09,Ofek_PNAS10,Halperin_PRB2011} interferometers allowed to study and control electron dephasing processes within ES and to estimate the electron phase coherence length~\cite{Neder_PRL2006,Roulleau_PRL2008,Huynh_PRL_2012}.
In particular, ref.~\cite{McClure_prl2009}  used interference checkerboard patterns generated in a Fabry-P\'erot geometry to determine the edge state velocity in the quantum Hall regime.  
In the other extreme, $\sigma<<2e^2/h$, Quantum Hall Islands (QHIs) created by patterned quantum dots~\cite{Altimiras_NatPhys09} or antidots~\cite{Kataoka_prl1999, Sim_pr2008,Maasilta_PRB98,Michael_phE2006,Goldman_sc95} lead to the observation of a fractional electric charge ~\cite{Goldman_sc95,Kou_prl2012,McClure_prl2012} and to the characterization of localized states' confining potential~\cite{Maasilta_PRB98,Sim_pr2008,Michael_phE2006}.

These pioneering experiments succeeded in evidencing electron interferences in model ES geometries, as well as signatures of charge tunneling through lithographically-patterned QHIs. Such phenomena can also occur in unpatterned two-dimensional electron systems: since the electron confining potential is never perfectly flat, electrons can be trapped in QHIs, pinned around potential "hills" and "dips". 
The contribution of these QHIs to electron transport depends on their coupling to propagating ES. It is therefore important to devise methods to explore electron transport at the local scale in such systems, and  new scanning probe techniques, derived from Atomic Force Microscopy (AFM) are now used for that purpose. 
The microscopic picture of electron localization was particularly investigated: compressibility measurements on quantum Hall localized states using a scanning single electron transistor revealed deviation from the single-particle picture~\cite{Ilani_Nature2004,Martin_NatPhys2009}. Moreover, localized states as well as the electrostatic potential confining the two-dimensional electron systems could be imaged by monitoring the charging of a moving `bubble' of electrons created in a two-dimensional electron system~\cite{Finkelstein_Science2000,Steele_PRL2005}.
Finally, Scanning Gate Microscopy (SGM)~\cite{Topinka_Science2000,Crook_PRB2000,Kicin_NJP2005,Schnez_NJP2011,Hackens_NatPhys2006,MartinsPRL2007,PalaPRB2008,PalaNano2009,Kicin_PRB2004,Aoki_PRB2005}, directly relates the microscopic structure of a quantum Hall system to its transport properties. 
At the starting point of the present work is the discovery that SGM allows to precisely locate QHIs connected to propagating ES through tunnel barriers, and to decrypt the complex magnetoresistance oscillations patterns to which they give rise~\cite{Hackens_NatComm2010,Paradiso_PRB2012}.

%-----------------------------------------------------------------------------------------------------------------------------------------------------------------------------------------------------------------
%\section{Summary}
Here, we go beyond SGM and apply Scanning Gate Spectroscopy (SGS)~\cite{Hackens_NatComm2010,Bleszynski-Jayich_prb2008} to access the discrete energy spectrum of an individual QHI pinned around a potential inhomogeneity. Unexpectedly, the QHI is located near one constriction of the quantum ring rather than around the central antidot.
In SGS, the microscope tip is used as a local gate whose bias tunes the energy levels inside the QHI that is coupled to propagating ES by tunneling.
Around filling factor $\nu$ = 6, the magnetoresistance of the quantum ring displays periodic oscillations. 
Thanks to SGM, we decrypt that they originate from Coulomb Blockade (CB) of electrons tunneling across a single ultra-small QHI.
The CB interpretation is confirmed by the presence of Coulomb diamonds in the SGS spectrum.
Importantly, lines parallel to the sides of Coulomb diamonds are associated with excited states arising from confinement.
Using their energy spacing, we estimate the gradient of the confining potential and the associated edge state velocity around the QHI.  

\section{Experimental setup and sample properties}

\begin{figure}
\begin{center}
\includegraphics[width=100 mm]{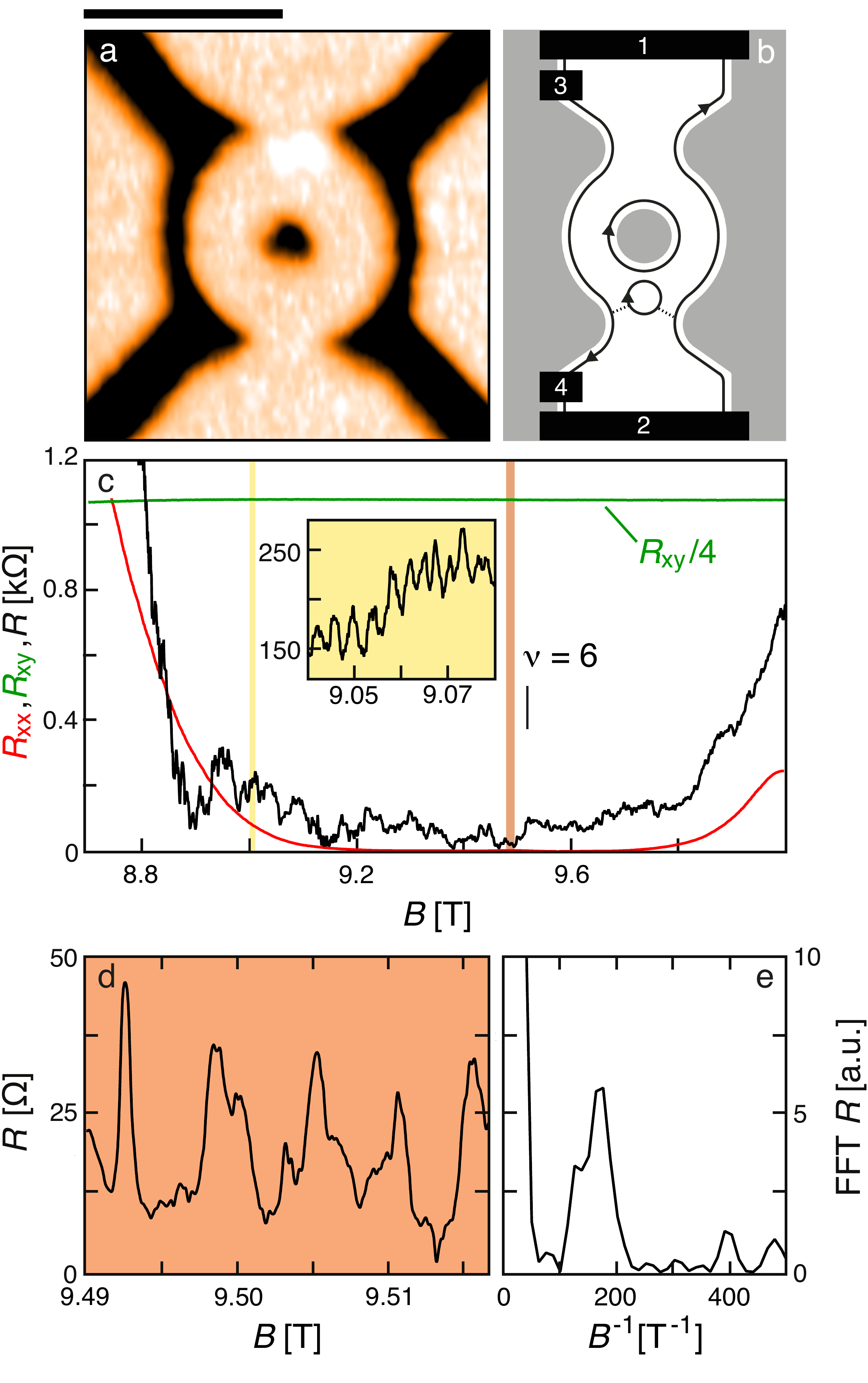}
\caption{(Color online)
(a) AFM topography of the quantum ring obtained at $B$~=~9.5~T and $T$~=~100~mK. The black bar represents 1~$\mu$m.
(b) Schematic representation of our device showing a QHI connecting opposite edge channels through tunnel barriers (dotted lines). Current carrying contacts (1-2) and voltage probes (3-4) allow resistance measurements. 
(c) Magnetoresistance of the quantum ring ($R$) together with the longitudinal resistance $R_{\mathrm{xx}}$, in red, and the transverse resistance $R_{\mathrm{xy}}$, in green, of a Hall bar a next to $\nu~=~6$. In the inset we zoom a set periodic oscillations corresponding to the yellow region.
(d) Close-up of the orange-shaded region in (c).
(e) Fast Fourier transform of magnetoresistance (d). 
}
\label{fig:fig1}
\end{center}
\end{figure}

Our experiments are carried out inside a $^{3}$He/$^{4}$He dilution refrigerator at temperature $T$~=~100~mK. 
The cryostat is equipped with a superconducting coil that provides a magnetic field up to 15/17~T. 
A home made cryogenic AFM is attached at the bottom of the mixing chamber. 
The movement of the tip is detected via a piezoelectric tuning fork to which a conductive cantilever is glued~\cite{Hackens_NatComm2010}. 
Using this setup, we perform measurements on a quantum ring patterned in an InGaAs/InAlAs heterostructure using e-beam lithography followed by wet etching. 
The two-dimensional electron gas is located 25~nm below the surface.
In figure~\ref{fig:fig1}(a) we show an AFM topography of our device obtained at $B$~=~9.5~T and $T$~=~100~mK, just before SGM measurements. 

As sketched in figure~\ref{fig:fig1}(b) the current is injected between ohmic contacts 1 and 2 and the voltage drop is measured between 3 and 4.  
Next to the ring, we patterned a Hall bar where we measure a low-$T$ electron density and mobility of $1.3 \times 10^{16}~\mathrm{m^{-2}}$ and $4~\mathrm{m^{2}/Vs}$, respectively. Additionally, two lateral gates visible in figure~\ref{fig:fig1}(a) were grounded during the experiments shown here.

With the AFM tip connected to the ground, and retracted, around $\nu~=~6$  (measured in the Hall bar), the magnetoresistance of the device, shown in figure~\ref{fig:fig1}(c), displays oscillations with various periods which correspond to different QHIs being successively "active"~\cite{Hackens_NatComm2010}, \emph{i.e.} tunnel-coupled to the propagating ES. A close up of a set of periodic oscillations is shown in the inset of figure~\ref{fig:fig1}(c). These magnetoresistance oscillations are not related to coherent electron interferences since their amplitude is inversely proportional to temperature which is consistent with pure Coulomb blockade~\cite{Hackens_NatComm2010}.
For the rest of our work, we focus on the orange-shaded region of figure~\ref{fig:fig1}(c),  zoomed in figure~\ref{fig:fig1}(d), which displays periodic peaks with $\Delta B =$~1/ 180~T~$\sim$~ 5.6~mT as evidenced by the fast Fourier transform displayed in figure~\ref{fig:fig1}(e). 
This magnetic field range was chosen after reviewing the full trace and picking one region where the magnetoresistance was characterized by one dominant frequency, and where the resistance minima were the closest to zero so that the edge state picture, i.e. essentially no backscattering of edge states by the constriction, was the simplest.
We understand these oscillations within a Coulomb-dominated model where electrons tunnel between propagating ES through a single QHI created around a potential inhomogeneity~\cite{Rosenow_PRL2007}. 
Intuitively, this model states that a variation in the magnetic field  creates an energy imbalance between the QHI and the propagating ES. 
This imbalance generates one CB-type oscillation per populated ES circling around the QHI per flux quantum $\phi_{\mathrm{0}}$. In this case, the device resistance oscillates with a period~\cite{Rosenow_PRL2007}:
\begin{equation}
	\Delta B= (\phi_{\mathrm{0}}/A)/N^*,
\label{rosenow}
\end{equation}
where $N^*$ is the number of filled ES around the QHI and $A$ is the QHI area. 
From equation~(\ref{rosenow}), figure~\ref{fig:fig1}(d), and assuming that $N^*=N=6$, where $N$ is the number of fully occupied ES in the bulk, we deduce that the QHI has a surface $A$ equivalent to that of a disk with a radius $r \simeq$~200~nm.
 
%------------------------------------------------------------------------------------------------------------------------------------------------------------------------------------------------------------------

\section{Imaging a QHI} 

\begin{figure}
\begin{center}
\includegraphics[width=100 mm]{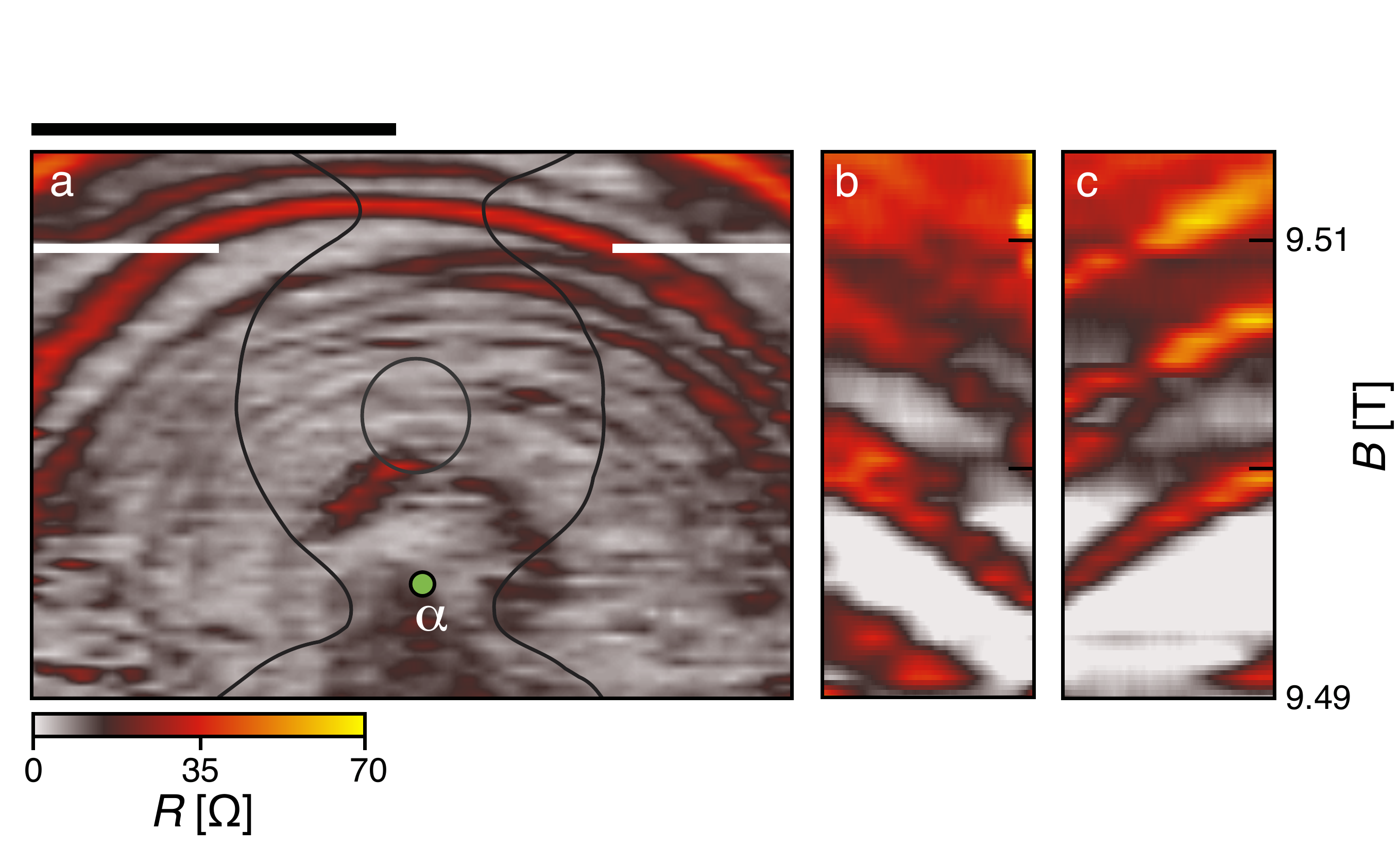}
\caption{(Color online)
(a) SGM map obtained at  $B$~=~9.5~T, $T$~=~100~mK, $D_{\mathrm{tip}}=50$ nm and $V_{\mathrm{tip}}$~=~2~V. Continuous lines indicate the position of the quantum ring whereas $\alpha$ marks the origin of concentric fringes. The black bar represents 1~$\mu$m.
(b-c) $B$-dependence of 500~nm long resistance profiles over the left/right regions marked with white lines in (a), respectively, with $V_{\mathrm{tip}}$~=~-1~V.
}
\label{fig:fig2}
\end{center}
\end{figure}

We now use SGM in order to reveal the position of the active QHI by tuning the local potential landscape.
It consists in scanning the polarized tip ($V_{\mathrm{tip}}$) along a plane parallel to the two-dimensional electron system at a distance $D_{\mathrm{tip}}=50$ nm, which roughly corresponds to the lateral extent of the tip-induced potential full width at half maximum \cite{Gildemeister_PRB2007}, while recording, at every point, the resistance of the device.
figure~\ref{fig:fig2}(a) shows a SGM map recorded at $B$~=~9.5~T, $V_{\mathrm{tip}}$~=~2~V and $T$~=~100~mK. 
The continuous black lines superimposed on the image indicate the lithographic edges of the quantum ring.
Concentric fringes decorate the SGM map and their center indicates the position of the active QHI (point $\alpha$ in figure~\ref{fig:fig2}(a)).
Indeed, approaching the polarized tip modifies the potential on the island.
It therefore gradually changes its area $A$ and hence the magnetic flux $\phi$ enclosed by circling ES.
Equivalently to the effect of $B$ described above, varying the tip-QHI distance also generates CB-type oscillations and isoresistance lines on figure~\ref{fig:fig2}(a) are indeed iso-$\phi$ lines.
Variations of the visibility of the fringes are observed when the tip approaches the QHI, a typical feature of CB experiments \cite{Kouwenhoven_AS1997,BennekerPRB1991}. 
Furthermore, the amplitude also varies along a single concentric fringe, most likely due to the variation of the tunneling strength through the QHI, which depends sharply on the potential landscape.
In figure~\ref{fig:fig2}(a), we chose $V_{\mathrm{tip}}$~=~2~V to make sure that $N^*=N=6$ during the entire scan. Otherwise, when approaching a negatively charged tip near the constrictions of the quantum ring, $N^*$ decreases and the resistance sharply increases \cite{Aoki_PRB2005}, which hinders the observation of CB-type fringes. Noteworthy, the same fringes are present in both negative and positive  $V_{\mathrm{tip}}$ as long as $N^*=N$, i.e. for both repulsive or attractive tip potential, respectively. Changing $V_{\mathrm{tip}}$, the fringes are simply shifted with respect to the position of the QHI.

The effects of both $B$ and tip-QHI distance along the white lines in the left and right sides of figure~\ref{fig:fig2}(a) are illustrated in figures~\ref{fig:fig2}(b) and (c), respectively (note that  figure~\ref{fig:fig2}(a) was obtained with positive $V_{\mathrm{tip}}$ while figure~\ref{fig:fig2}(b) and (c) with negative $V_{\mathrm{tip}}$).
Along the $B$-axis, clear oscillations are visible with a period around $\Delta B\sim 6$~mT in agreement with the magnetoresistance in figure~\ref{fig:fig1}(d).
For the case of a potential hill, resistance peaks should follow positive $dV_{\mathrm{tip}}/dB$ slopes, i.e. decreasing $B$ should be compensated by a tip voltage decrease in order to keep the same magnetic flux through the QHI~\cite{Zhang_PRB09,Ofek_PNAS10,Halperin_PRB2011}.
In our case, approaching the negatively charged tip has the same effect as decreasing $V_{\mathrm{tip}}$ as it corresponds to an increase of the QHI area. The slope of the fringes in figures~\ref{fig:fig2}(b-c) therefore supports the CB picture. 

Finally, the ES model states that a complete reflection of $K$ edge channels by a quantum point contact gives rise to a resistance shift of $\frac{h}{e^2}\left (\frac{1}{N-K}-\frac{1}{N} \right )$~\cite{Aoki_PRB2005, Buttiker_PRB1988}.
Using this argument~\cite{Buttiker_PRB1988} with $N$~=~6 and taking into account that the amplitude of the oscillations is around $40\Omega$, we deduce $K$~=~0.05, 
meaning that the peak conductance across the QHI is about 0.05 $e^2/h$, also consistent with the tunneling regime required for CB. 
This also means that most of the current flows in the transmitted ES and only a small fraction is used to probe the QHI, in contrast with typical CB experiments where all the current flows through the quantum dot.
 
\section{Local spectroscopy of the QHI}

\begin{figure*}
\begin{center}
\includegraphics[width=100 mm]{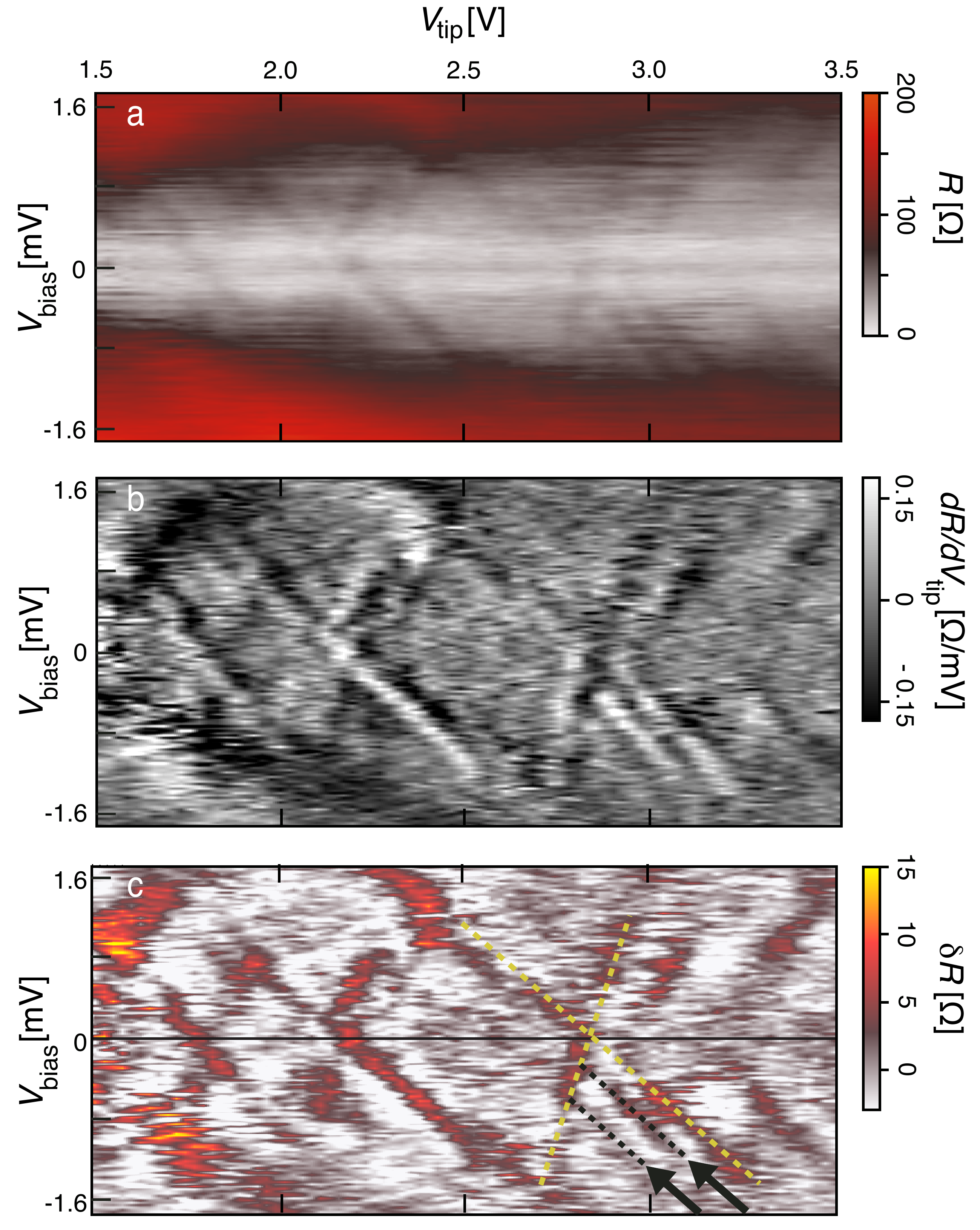}
\caption{(Color online)
(a) SGS measured when placing the tip at $\alpha$, in figure~\ref{fig:fig2}(a). 
(b) Numerical derivative $dR/dV_{\mathrm{tip}}$ of (a) in the $V_{\mathrm{tip}}-V_{\mathrm{bias}}$ plane. Note that each resonance in the differential resistance appears as a black and white double line.
(c)~High-pass-filter version of (a), where two yellow dashed lines indicate the transition between Coulomb-blockaded regions. Black dashed lines highlight Coulomb excited states.
}
\label{fig:fig3}
\end{center}
\end{figure*}

A deeper understanding of the QHI's electronic structure can be reached using SGS: positioning the tip at point $\alpha$, both $V_{\mathrm{tip}}$ and the current $I$ through the device are swept. 
The bias current can be converted to a voltage bias $V_{\mathrm{bias}}$ across the QHI thanks to the quantized Hall resistance: $V_{\mathrm{bias}}=h/(e^2 N)I$.
The plot of the differential resistance $R$ in the ($V_{\mathrm{tip}}$, $V_{\mathrm{bias}}$) plane is drawn in figure~\ref{fig:fig3}(a), where a set of narrow straight lines is superimposed on a slowly-varying background. 
We associate this background to the breakdown of the QH regime as a large current flows through the device~\cite{Nachtwei_PhyE1999}.
Straight lines are more apparent in figure~\ref{fig:fig3}(b), where the numerical derivative $dR/dV_{\mathrm{tip}}$ of figure~\ref{fig:fig3}(a) is plotted, as well as in figure~\ref{fig:fig3}(c), where a smooth background was subtracted from the raw $R$ vs $[V_{\mathrm{tip}},V_{\mathrm{bias}}]$ data to obtain $\delta R$.
Consistently with previous observations, crossing lines (two of them are highlighted with yellow dashed lines) correspond to transitions between Coulomb blockaded regions as energy levels of the QHI enter in the bias window defined by the source and drain electro-chemical potentials.
The slopes are determined by the various capacitances $C_\Sigma$ of the system. Crossing lines form Coulomb diamonds analog to those measured in closed quantum dots at $B$~=~0~T~\cite{Kouwenhoven_AS1997}. 

Parallel to the borders of the Coulomb diamonds, additional lines are visible, highlighted with black dashed lines in figure~\ref{fig:fig3}(c). Noteworthy, these lines do not enter the adjacent Coulomb diamond, a characteristic signature of excited states~\cite{Reimann_RMP_2002}.
The energy gap between these excited states is $\Delta E\simeq$ 380~$\mu$eV, much smaller than the Landau level spacing, $\Delta E_{\mathrm{LL}} \simeq 26.8$~meV,  and the Zeeman splitting, $\Delta E_{\mathrm{z}}\simeq1.65$~meV, in our system around $B= 9.5$~T. 
These values are determined making use of  $\Delta E_{\mathrm{LL}}=\hbar e B/m^*$, where $e$ is the electron charge, $m^* = 0.041~m_{\mathrm{e}}$ is the electron effective mass in our  two-dimensional electron system with respect to the free electron mass $m_{\mathrm{e}}$ \cite{Kotera_PhysE2001}, and $\Delta E_{\mathrm{z}}=\vert g \vert \mu_{\mathrm{\beta}} B$, where  $\mu_{\mathrm{\beta}}$ is the Bohr magneton and $g\simeq 3$ is the Land\'{e} g-factor in the heterostructure \cite{Nitta_APL2003}.
The QHI is thus in the quantum limit of CB defined by $k_{B}T<< \Delta E << e^2/C_\Sigma \sim1.6$ meV obtained from figure~\ref{fig:fig3}(b)~\cite{Kouwenhoven_AS1997}, which justifies the $1/T$ temperature dependence found previously in~\cite{Hackens_NatComm2010}.
Assuming that the Coulomb excited states emerge as a consequence of circular confinement, we can determine an average of the derivative of the confining potential $dU/dr$ using the relation~\cite{Kataoka_prl1999, Sim_pr2008,Michael_phE2006}:
\begin{equation}
	\Delta E= \hbar v/r =\vert dU/dr \vert \hbar/(e~rB),
\label{Sim}
\end{equation}
where $v$ is the ES velocity in the QHI.
Taking into account that $r \simeq$~200~nm, we obtain $\vert dU/dr \vert  \simeq$~1 meV/nm for the QHI formed around a potential inhomogeneity. 
In QHIs created around antidots etched in GaAs/AlGaAs heterostructures~\cite{Kataoka_prl1999,Maasilta_PRB98,Michael_phE2006}, which are known to generate soft-wall potentials, $\vert dU/dr \vert$ was found to be around 20~$\mu$eV/nm~\cite{Maasilta_PRB98}, and data in~\cite{Kataoka_prl1999,Michael_phE2006} give  $\vert dU/dr \vert \simeq$~50~$\mu$eV/nm.
In the case of a GaAs quantum Hall Fabri-P{\'e}rot interferometer~\cite{McClure_prl2009}, in the high-magnetic-field limit, $\vert dU/dr\vert $ was found to be 80~$\mu$eV/nm.
These results are summarized in table~(\ref{tab:table1}), from which we conclude that $\vert dU/dr \vert$ scales with the carrier density of the  two-dimensional electron system~\cite{Eksi _prb2007}: $1 \times 10^{15}~\mathrm{m^{-2}}$~\cite{Maasilta_PRB98}, $2.6 \times 10^{15}~\mathrm{m^{-2}}$~\cite{Kataoka_prl1999,Michael_phE2006},  $2.7 \times 10^{15}~\mathrm{m^{-2}}$~\cite{McClure_prl2009} and $1.4 \times 10^{16}~\mathrm{m^{-2}}$ in the present sample.
In the same framework, using equation~(\ref{Sim}), we extract $v = 9.8 \times 10^4$ m/s, which is close to the lower limit of results in~\cite{McClure_prl2009}. As $v$ decreases with increasing $B$, this data copes with the larger values reported in~\cite{McClure_prl2009} at lower $B$.
The numerical values determined here for $v$ and $\vert dU/dr\vert $ constitute upper bounds as the shape of the QHI is assumed to be circular.

\begin{table}[t]
\centering
\begin{tabular}{ |c|c|c| }
  \hline
  Reference & Carrier density (${m^{-2}}$)& $\vert dU/dr\vert $ ($\mu$eV/nm)\\
    \hline  \hline  
\cite{Maasilta_PRB98}& $1 \times 10^{15}$ & 20 \\
  \hline 
  \cite{Kataoka_prl1999,Michael_phE2006}& $2.6 \times 10^{15}$ & 50\\
  \hline
  \cite{McClure_prl2009} & $2.7 \times 10^{15}$  & 80\\ 
  \hline
this work  &$1.4 \times 10^{16}$ & 1000\\
  \hline
\end{tabular}
\caption{Comparison between $dU/dr$ calculated from data in the literature and extracted in this work as a function of the carrier density in the 2DES.}
\label{tab:table1}
\end{table}

\section{Conclusion}
In conclusion, we have used SGM to locate an individual quantum Hall island and directly probed its excited states arising from confinement with SGS.
We were able to confirm that Coulomb blockade governs transport and to characterize the confining potential of the QHI. Both the slope of the confining potential and the edge state velocity were estimated. 
The combination of SGM and SGS techniques is therefore extremely useful to unveil the microscopic nature of charge transport inside complex nanodevices even in the quantum Hall regime, just like the combination of scanning tunneling microscopy and scanning tunneling spectroscopy was key to understand the electronic local density of states of surface nanostructures.
Another advantage of the presence of the tip is the possibility to tune  \emph{in situ} the local electrostatic environment of the nano-device by depositing charges on its surface~\cite{Crook _Nature2003}. This way, one could envision to induce QHIs with controlled geometries.

\section*{Acknowledgements}

F.M. and B.H. acknowledge support from Belgian FRS-FNRS.
This work has been supported by FRFC grant no. 2.45003.12 and FNRS grant no. 1.5.044.07.F and by the Belgian Science Policy (Interuniversity Attraction Pole Program IAP-6/42).
This work has also been supported by the PNANO 2007 program of the Agence Nationale de la Recherche (MICATEC project). V.B. acknowledges the award of a Chaire d'excellence by the Nanoscience Foundation in Grenoble.
\bibliographystyle{prsty}

\end{document}